\documentclass{aastex}
\usepackage{spr-astr-addons}
\usepackage{url}
\usepackage{graphicx}
\usepackage{epstopdf}
\usepackage[fleqn]{amsmath}
\RequirePackage{color}

\usepackage[colorlinks=true,linkcolor=blue,citecolor=blue,urlcolor=blue]{hyperref}%

\begin{document}

\title{Whistler instability stimulated by the suprathermal electrons present in space plasmas}

\slugcomment{Not to appear in Nonlearned J., 45.}
\shorttitle{Whistler instability}
\shortauthors{Shaaban et al.}

\author{M.~Lazar\altaffilmark{1,2}} \and \author{R.~A.~L\'opez \altaffilmark{1}} \and \author{S.~M.~Shaaban \altaffilmark{1,3}} \and \author{S.~Poedts \altaffilmark{1}} \and \author{H.~Fichtner\altaffilmark{2}}

\altaffiltext{1}{Centre for Mathematical Plasma Astrophysics, Celestijnenlaan 200B, B-3001 Leuven, Belgium.} \email{shaaban.mohammed@student.kuleuven.be}
\altaffiltext{2}{Institut f\"ur Theoretische Physik, Lehrstuhl IV: Weltraum- und Astrophysik, Ruhr-Universit\"at Bochum, D-44780 Bochum, Germany.}
\altaffiltext{3}{Theoretical Physics Research Group, Physics Department, Faculty of Science, Mansoura University, 35516, Egypt.}

\begin{abstract}
In the absence of efficient collisions, deviations from thermal equilibrium of plasma particle 
distributions are controlled by the self-generated instabilities. The whistler instability
is a notorious example, usually responsible for the regulation of electron temperature 
anisotropy $A = T_{\perp}/T_\parallel>$ (with $\perp, \parallel$ respective to the magnetic 
field direction) observed in space plasmas, e.g., solar wind and planetary magnetospheres. 
Suprathermal electrons present in these environments change the plasma 
dispersion and stability properties, with expected consequences on the kinetic instabilities 
and the resulting fluctuations, which, in turn, scatter the electrons and reduce 
their anisotropy. In order to capture these mutual effects we use a quasilinear kinetic 
approach and PIC simulations, which provide a comprehensive characterization of 
the whistler instability under the influence of suprathermal electrons. Analysis is performed 
for a large variety of plasma conditions, ranging from low-beta plasmas encountered in 
outer corona or planetary magnetospheres to a high-beta solar wind characteristic
to large heliospheric distances. Enhanced by the suprathermal electrons, whistler fluctuations 
stimulate the relaxation of temperature anisotropy, and this influence of suprathermals 
increases with plasma beta parameter.
\end{abstract}


\section{Introduction and motivation}

Whistler instability is known as the instability of electromagnetic electron-cyclotron 
(EMEC) modes, driven unstable by the free energy of anisotropic electrons, in 
particular, those with anisotropic temperatures $T_\perp > T_\parallel$, (where $\perp, 
\parallel$ denote directions respective to the local magnetic field lines) \citep{Kennel1966}. 
This instability has given rise to a particular interest in space plasmas, where it is 
often invoked to explain, for instance, the electromagnetic fluctuations detected at 
kinetic scales in the solar wind and terrestrial magnetosphere \citep{Wilson2009, 
Wilson2013, Kellogg2011} and, implicitly, the role played by these fluctuations in the 
relaxation of temperature anisotropies \citep{Stverak2008, Schriver2010, An2017, Kim2017}. 
The interested readers may also consult \cite{Stenzel1999} for a review of the early 
observations of whistlers in the solar wind and planetary magnetospheres.

The knowledge of the velocity distribution of electrons is particularly important
for a realistic characterization of whistler instability and its implications. The in-situ
measurements in space plasmas across the heliosphere reveal non-thermal distributions 
with non-Maxwellian suprathermal tails \citep{Vasyliunas1968, Maksimovic1997, Pierrard2010}.
In theory commonly adopted are the idealized (bi-)Maxwellian models \citep{Gary1993, Yoon2017}, 
which can reproduce only the low-energy core of the observed distributions. The high density 
of the core population (with more than 90\% of total number of particles) is invoked as a 
justification, although an important amount of heat and kinetic energy is transported by the 
suprathermal populations \citep{Pierrard2010,Lazar2015EMEC}. Suprathermal electrons are ubiquitous in the 
solar wind enhancing the high-energy tails of the observed distributions and highly contrasting 
with the (bi-)Maxwellian core \citep{Maksimovic1997}. The role of 
these hotter electrons in the excitation of whistler instability has been anticipated in the 
original study of \cite{Kennel1966}, though these authors have treated only waves 
resonant with electrons on the high-energy tail, where the number of resonant particles, and 
therefore the growth rate, is small. In the present paper we present a complete analysis of 
the whistler instability resulting from the interplay of both thermal and suprathermal 
electrons, and outline the effects of suprathermal electrons on the enhanced fluctuations 
and the relaxation of temperature anisotropy.

As an intimate component of the non-equilibrium plasmas, suprathermal populations are 
expected to be an important source of free energy, triggering various plasma processes 
\citep{Hapgood2011}. If we refer to kinetic instabilities driven by temperature anisotropies, 
many of the existing studies involving a bi-Kappa representation fail to describe 
the effects of suprathermal populations. Thus, contrary to expectations, suprathermals 
appear to have an inhibiting effect, and, particularly for the whistler instability, 
stimulation has been found only for marginal conditions, i.e., for very low anisotropies 
$1 \lesssim T_\perp/T_\parallel <2$ \citep{Mace2010, LazarAA2013, LazarJGR2018}. However, 
in our study we apply a straightforward comparative analysis recently proposed to outline 
suprathermal populations and their implications by contrasting the observed Kappa distribution 
with a quasi-thermal core population \citep{Lazar2015Destabilizing, LazarAA2016}. 
The anisotropic electrons are described by a single or global bi-Kappa 
(or bi-Lorentzian) distribution 
function \citep{Summers1991}, which is nearly bi-Maxwellian at low energies reproducing the core, and 
decreases as a power-law at higher energies \citep{Pierrard2010,Lazar2015Destabilizing, LazarAA2017}. 
Dual models involving Kappa or bi-Kappa functions to reproduce only suprathermal (or halo) 
population in combination with a bi-Maxwellian to describe the core \citep{Maksimovic2005, 
Stverak2008, LazarAA2017, Wilson2019}, may give better fits to the observations than a single 
bi-Kappa incorporating both the core and halo populations \citep{LazarAA2017}. However, a single,
or global bi-Kappa representation is more often invoked in theoretical analyses due to a 
reduced number of parameters involved \citep{Mace2010, LazarAA2013,Lazar2015Destabilizing}, and 
has a particular relevance for describing the effects of suprathermal populations 
\citep{Lazar2015Destabilizing, LazarAA2016}. It may also be motivated observationally, by 
the solar wind electron data collected by spacecraft missions from different heliospheric distances 
\citep{Stverak2008}, which are dominated by the core (subscript c) and halo (subscript h) 
populations with similar anisotropies, either both with $A_{c,h} > 1$ or both with $A_{c,h} < 1$ 
\citep{Pierrard2016}. These states with correlated anisotropies of the core and halo populations 
are the most unstable and most relevant for instabilities (of interest for us being those with $A_c 
\simeq A_h > 1$), and for the sake of simplicity and generality, we assume them well described by 
a single bi-Kappa, see also Appendix A. A general analysis of suprathermal populations and their effects becomes indeed
straightforward, by contrasting with the results obtained for quasi-thermal core well approached
by the bi-Maxwellian limit $\kappa \to \infty$, see Appendix A and \cite{Lazar2015Destabilizing,
LazarAA2016,Lazar2018Spont}. 

Such a direct comparison leads to a systematic stimulation of linear growth rates in 
the presence of suprathermal populations, and not only for the whistler instability 
\citep{Lazar2015Destabilizing, Vinas2015, ShaabanASS2016, ShaabanASS2017, LazarMNRAS2017, 
ShaabanJGRA2018, ShaabanMNRAS2019}. In order to test these predictions 
from linear theory here we propose a comprehensive study of whistler instability 
employing quasilinear theory (section 2) and particle-in-cell (PIC) simulations 
(section 3). The effects of suprathermal electrons are evaluated for a wide range of values 
of the plasma beta parameter, covering heliospheric plasma conditions specific to the solar 
wind and planetary magnetospheres.

\begin{figure*}[h!t!]
   \centering
\includegraphics[width=0.3\textwidth]{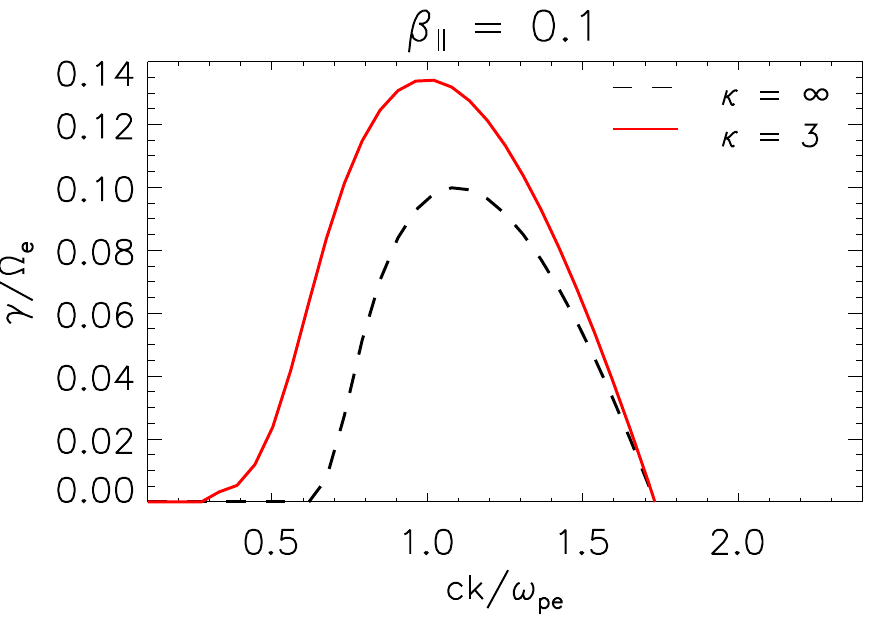} \includegraphics[width=0.3\textwidth]{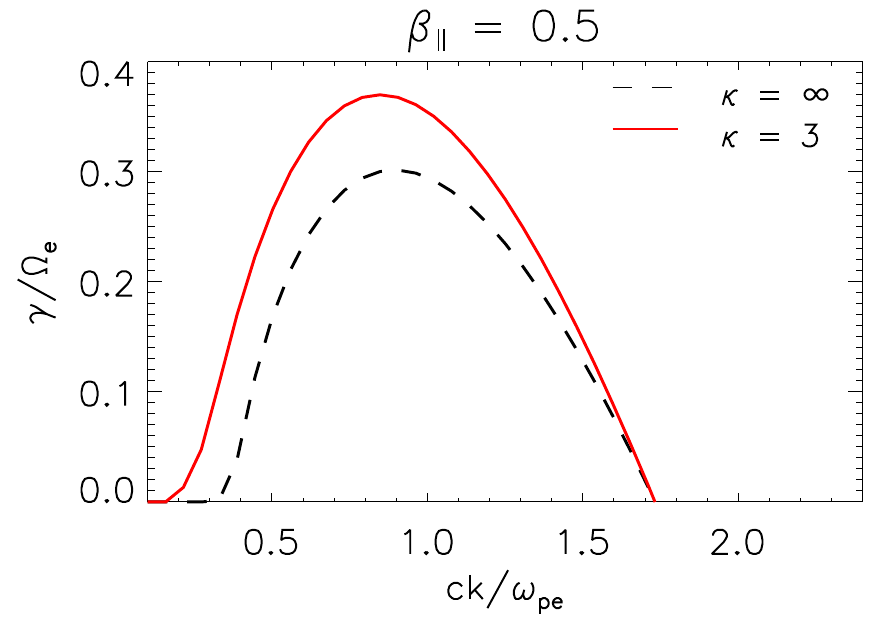}
\includegraphics[width=0.3\textwidth]{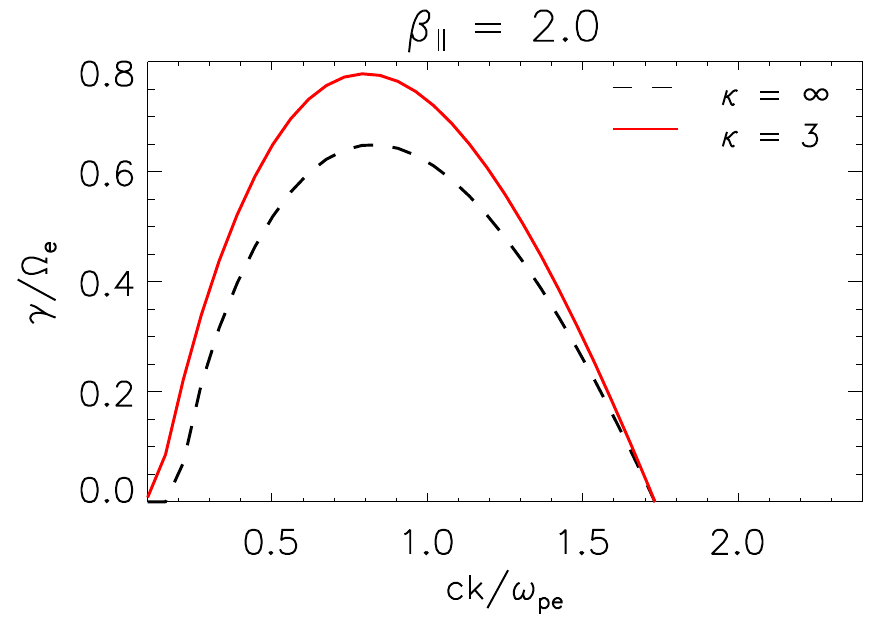}\\
\includegraphics[width=0.3\textwidth]{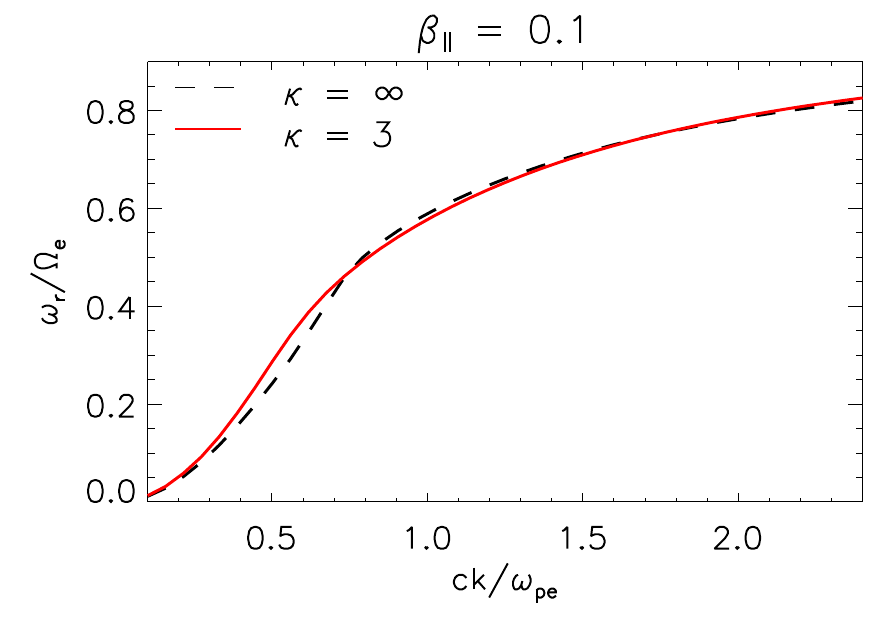} \includegraphics[width=0.3\textwidth]{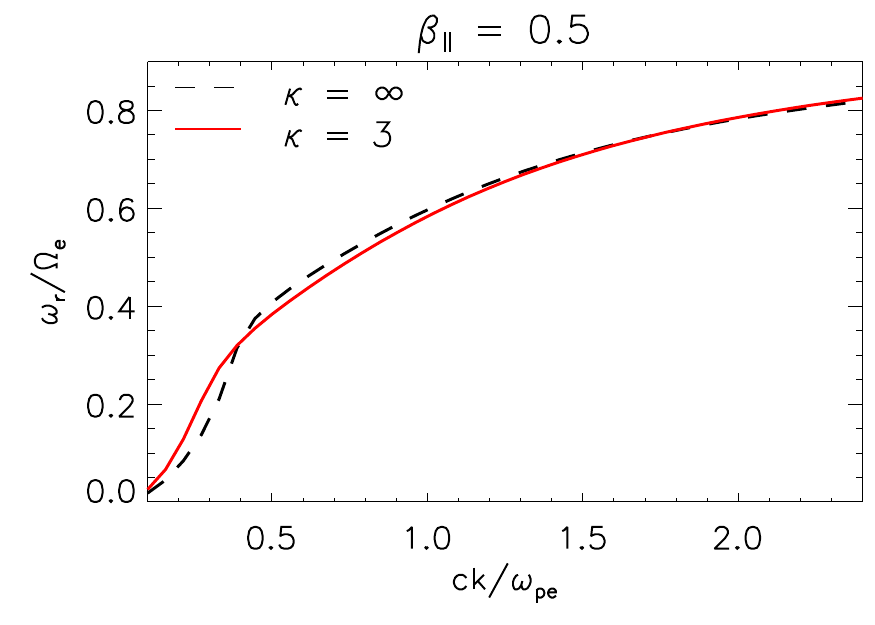}
\includegraphics[width=0.3\textwidth]{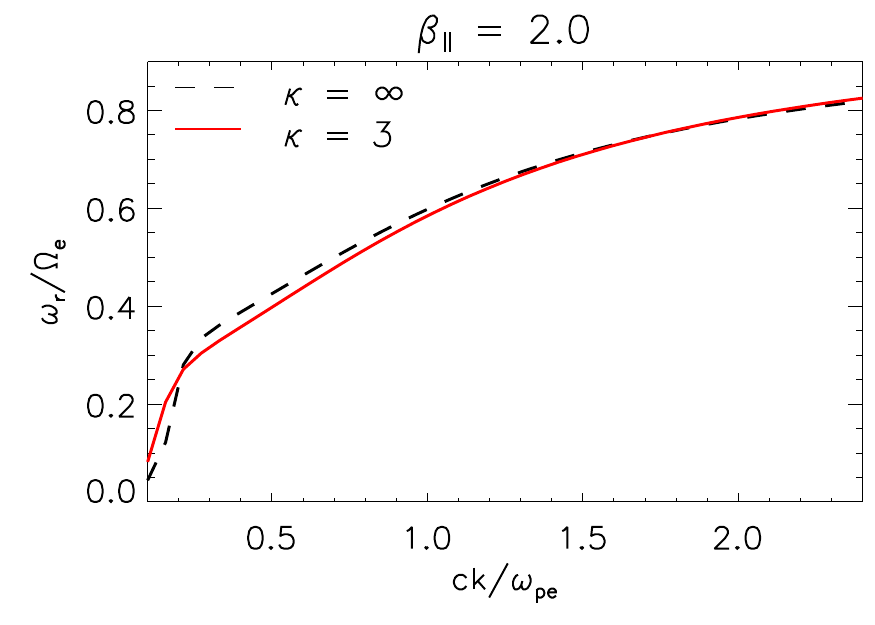}\\
\includegraphics[width=0.3\textwidth]{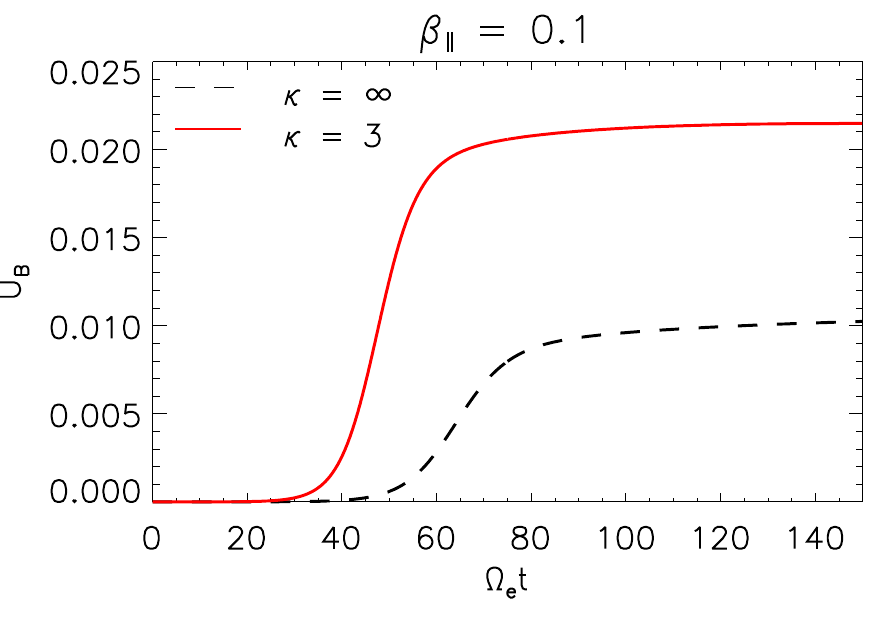} \includegraphics[width=0.3\textwidth]{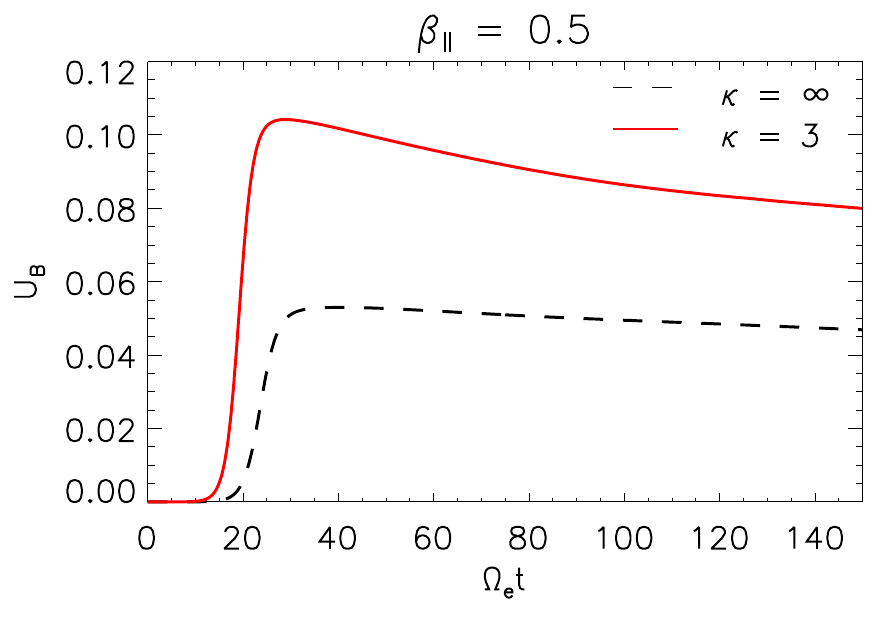}
\includegraphics[width=0.3\textwidth]{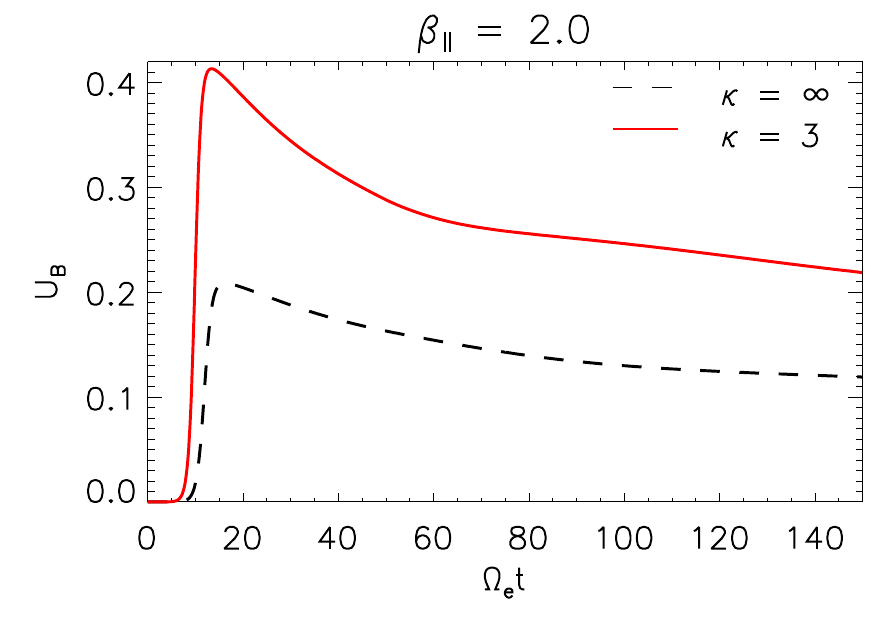}\\
\includegraphics[width=0.3\textwidth]{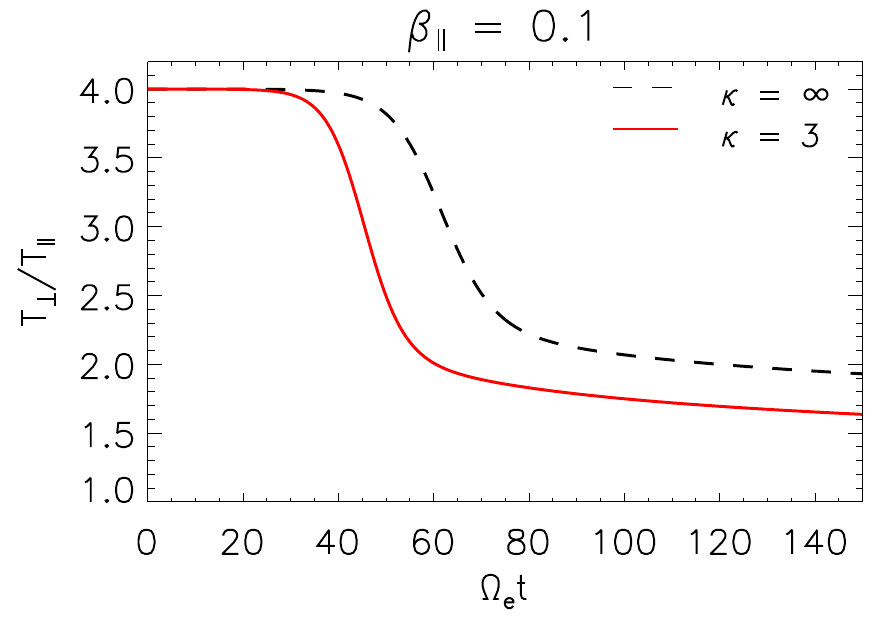} \includegraphics[width=0.3\textwidth]{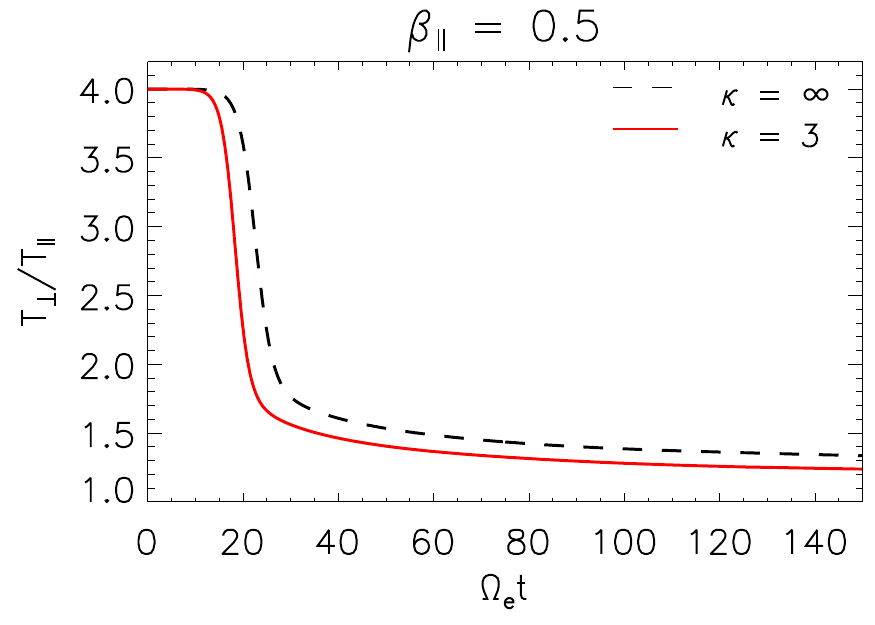}
\includegraphics[width=0.3\textwidth]{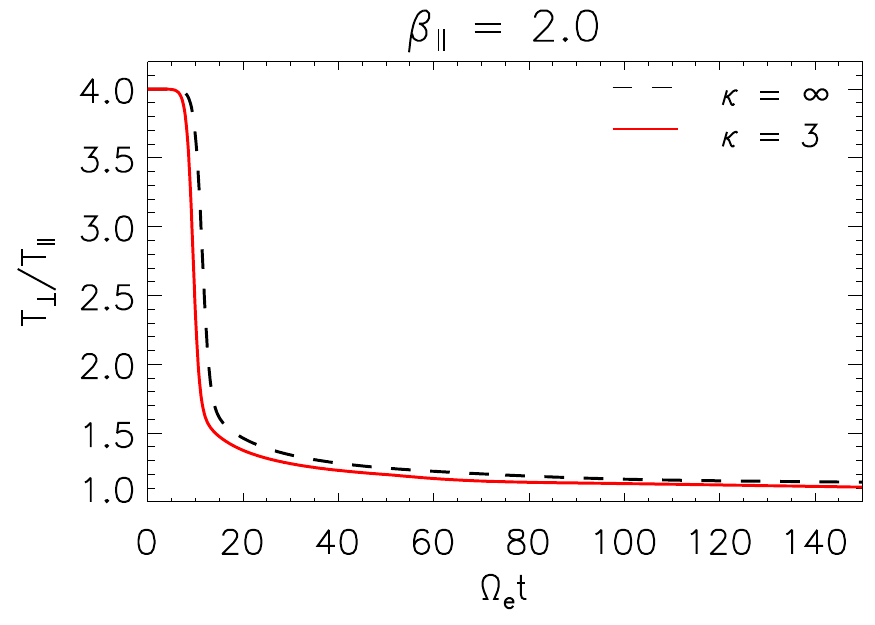}
\caption{Systematic stimulation of the whistler instability (growth rate $\gamma$, 
wave-frequency $\omega_r$, magnetic power $U_B$) and subsequent relaxation of 
temperature anisotropy ($T_\perp / T_\parallel$) in the presence of suprathermals: 
three distinct cases corresponding to $\beta_{c,\parallel} (0) = \beta_\parallel (0) = 
0.1, 0.5, 2.0$, see discussion in the text.} \label{f1}
\end{figure*}

\begin{figure*}[h!]
   \centering
\includegraphics[width=0.32\textwidth]{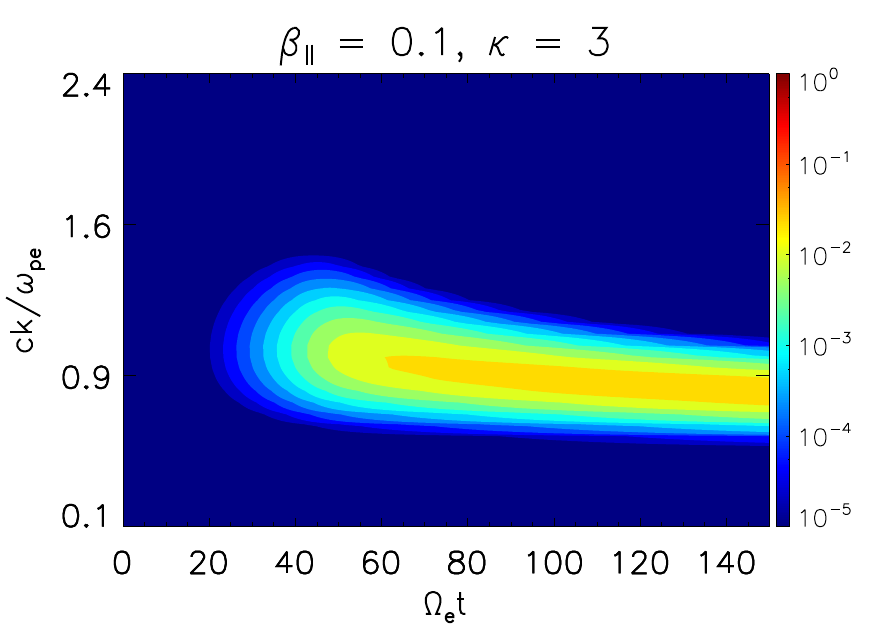} \includegraphics[width=0.32\textwidth]{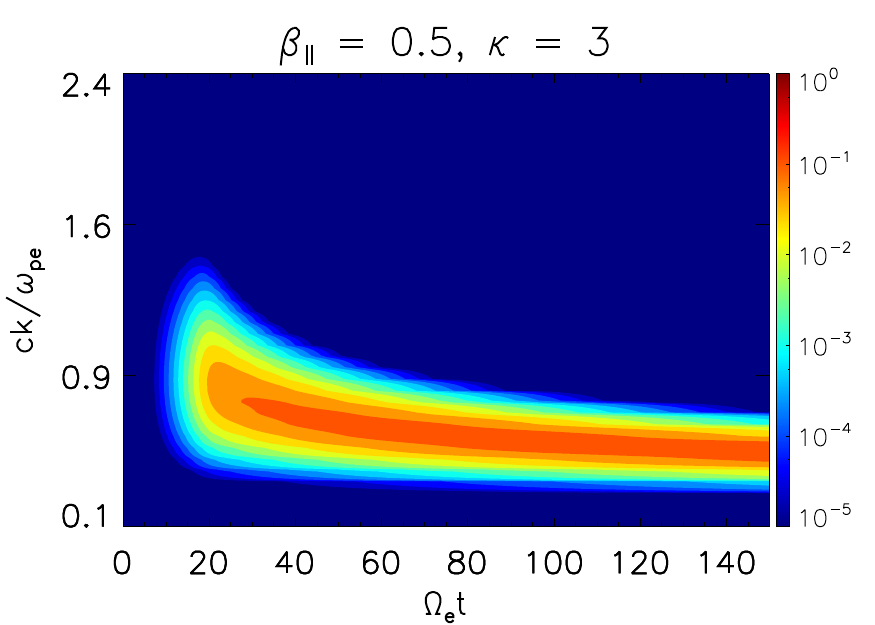}
\includegraphics[width=0.32\textwidth]{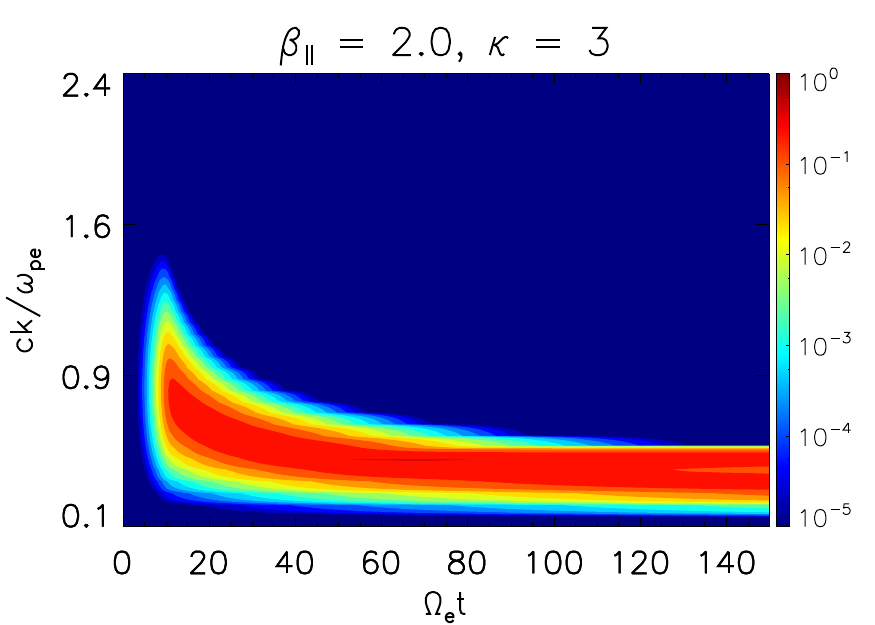}\\
\includegraphics[width=0.32\textwidth]{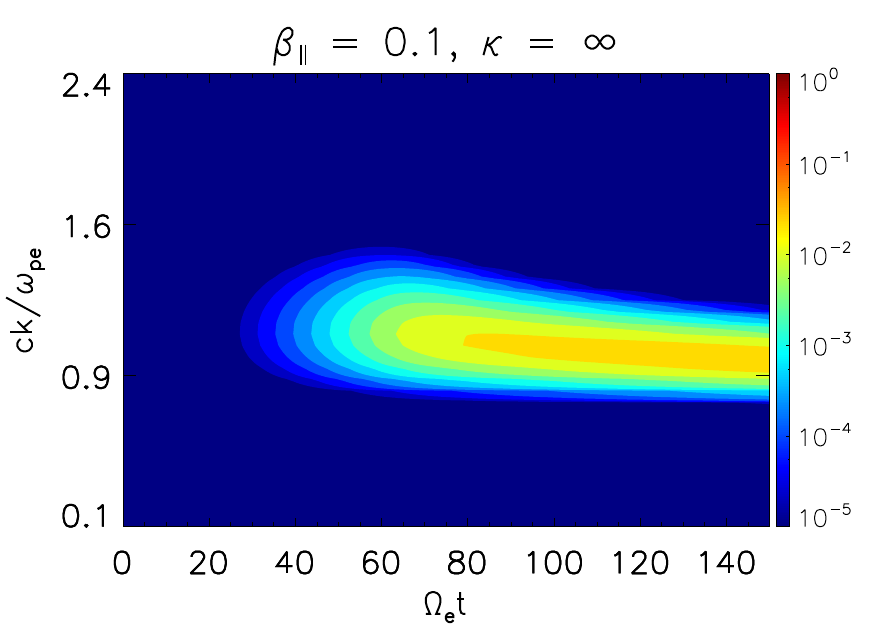} \includegraphics[width=0.32\textwidth]{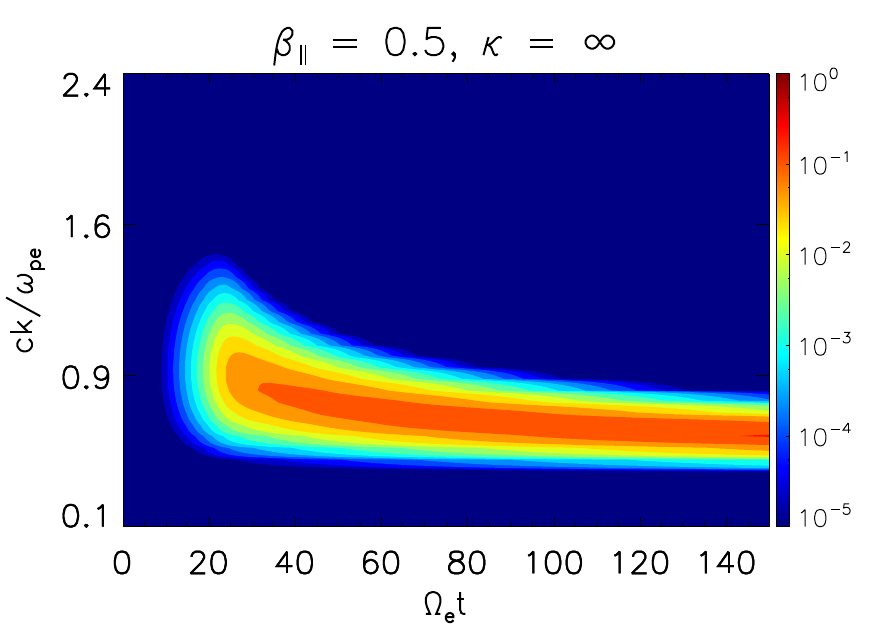}
\includegraphics[width=0.32\textwidth]{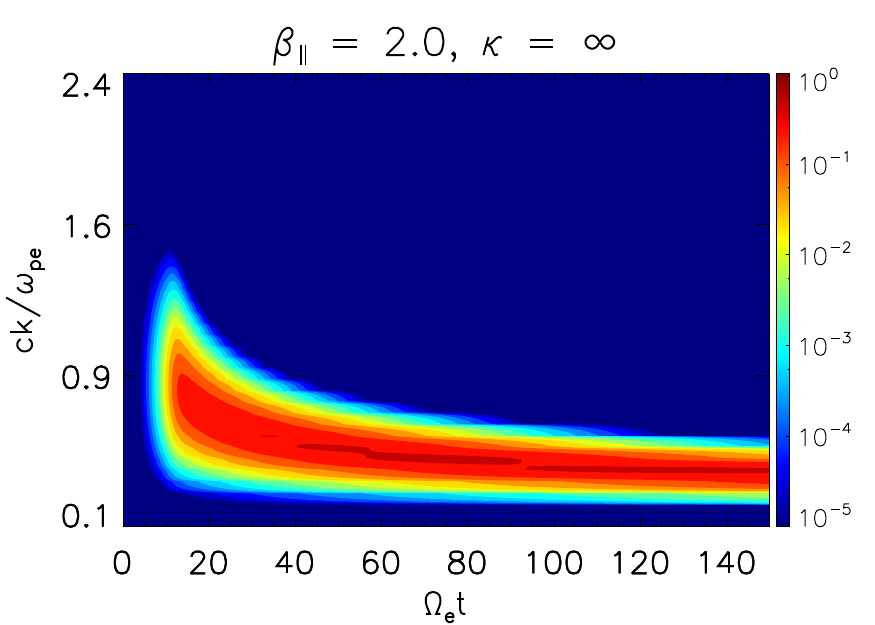}
\caption{Temporal evolution of the wave-number spectra of the color-coded (normalized) magnetic 
wave power, $\delta B^2(k)/B_0^2$, which is enhanced by the suprathermal electrons (upper panels) 
at early times.}\label{f2}
\end{figure*}

\begin{figure}[h!]
   \centering
\includegraphics[width=0.45\textwidth]{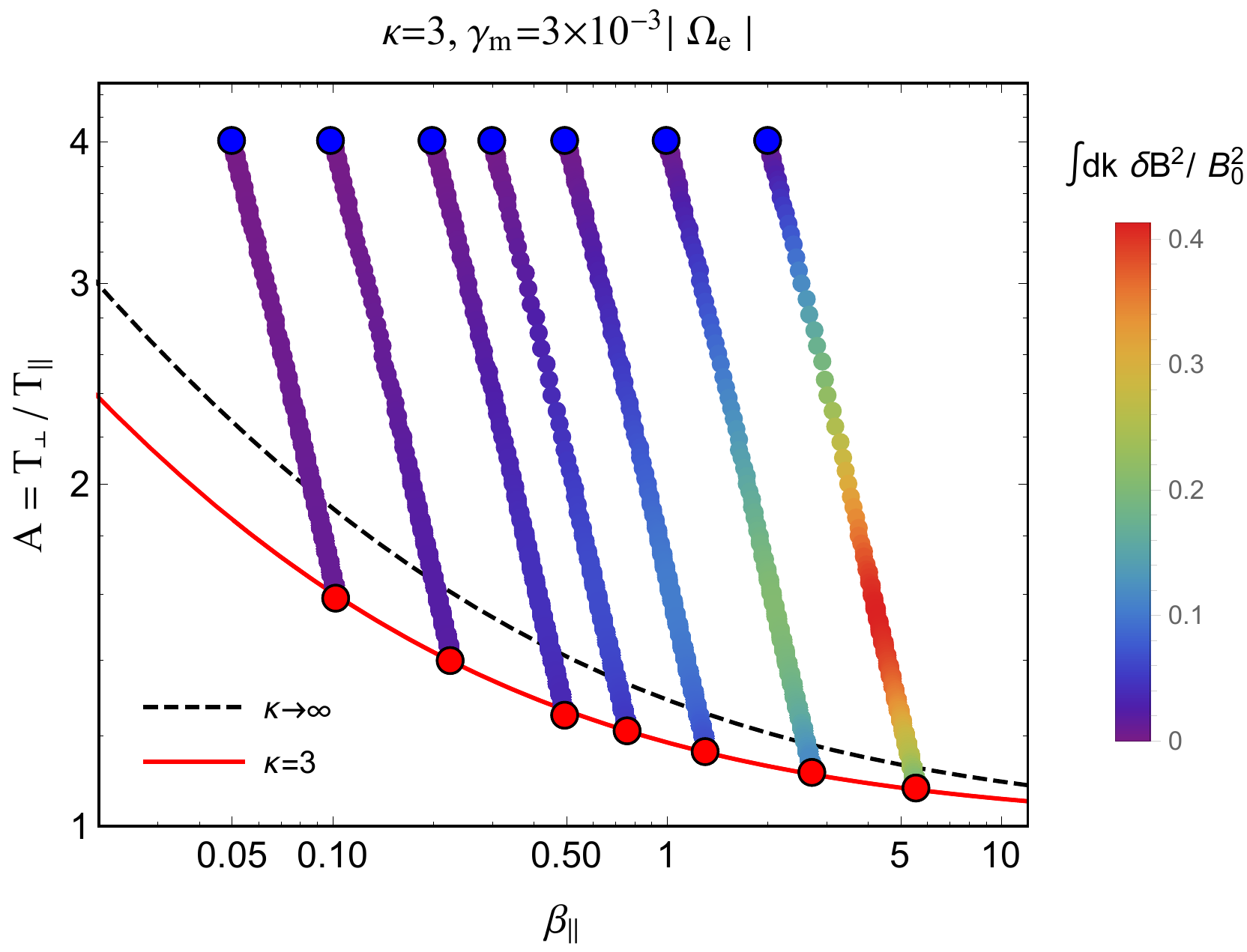} \includegraphics[width=0.45\textwidth]{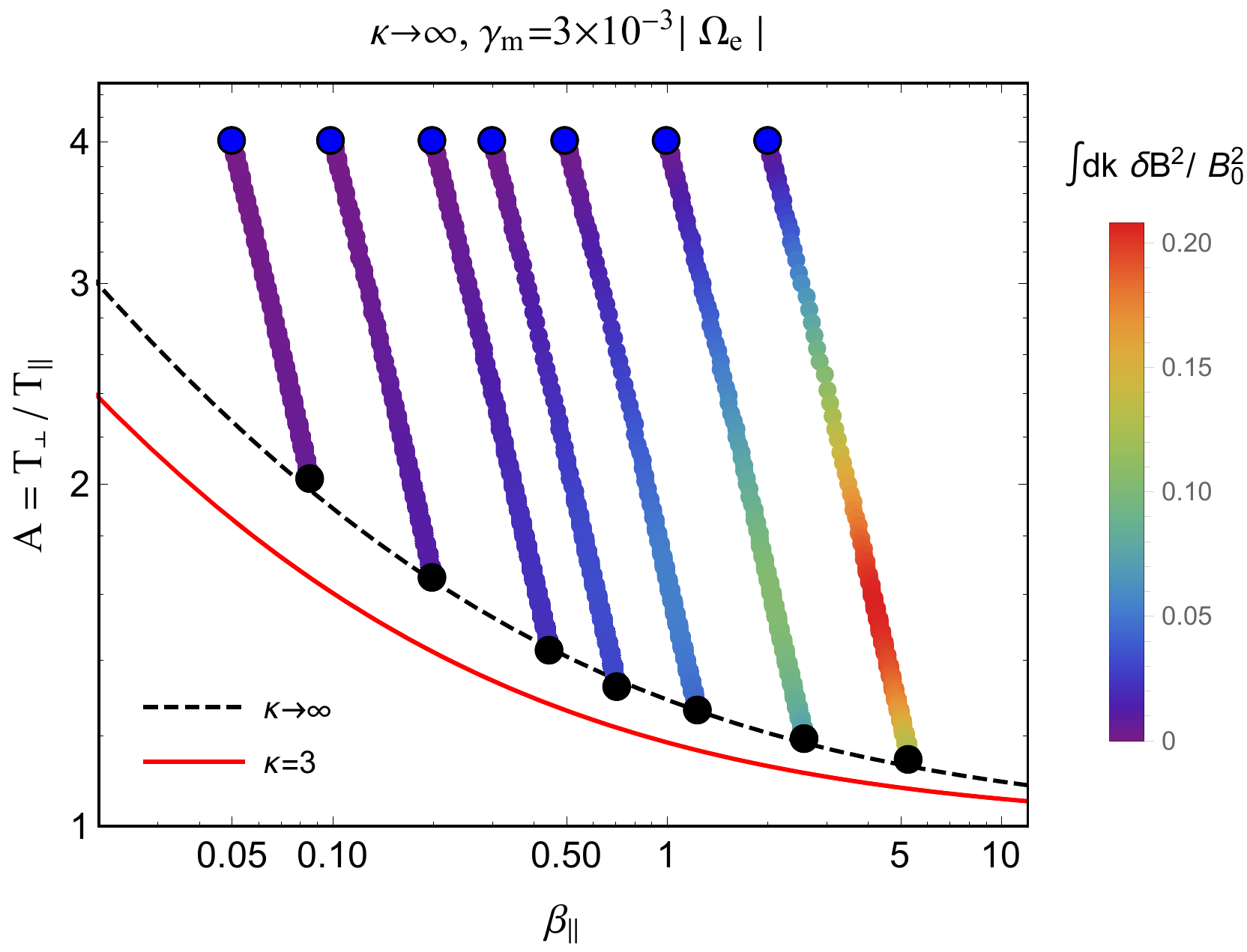}
\caption{QL dynamical paths end-up on the instability thresholds, which decrease in the 
presence of suprathermal electrons (red lines).}\label{f3}
\end{figure}

\section{Quasilinear approach}     

We assume a sufficiently homogeneous and collisionless plasma of bi-Kappa
distributed electrons (subscript $e$) and Maxwellian isotropic protons 
(subscript $p$), see Appendix~A. Whistler modes are described by the linear 
(instantaneous) dispersion relation 
\begin{align} \label{e1}
\tilde{k}^2= A -1 &+ \frac{A~\tilde{\omega} - A -1}{\tilde{k}\sqrt{\beta_{e,\parallel}}} 
Z_\kappa \left(\frac{\tilde{\omega}-1}{\tilde{k} \sqrt{\beta_{e,\parallel}}}\right)\nonumber\\
&+\frac{\tilde{\omega}}{\tilde{k}~\sqrt{\mu~\beta_p}}Z_p\left(\frac{\mu~\tilde{\omega}+1}
{\tilde{k}\sqrt{\mu~\beta_p}}\right),
\end{align}
where $\tilde{\omega}= \omega/|\Omega_e|$ is the normalized wave frequency $\omega = 
\omega_r + i \gamma$, $\tilde{k}=kc/\omega_{p,e}$ is the normalized wave-number $k$, 
$c$ is the speed of light in vacuum, $\omega_{p, e}=\sqrt{4\pi n_0 e^2/m_e}$ and $\Omega_e
= e B_0/(mc)$ are, respectively, the plasma and cyclotron frequencies of electrons, 
$\mu= m_p/m_e= 1836$ is the proton--electron mass ratio. Plasma beta parameters 
$\beta_{\parallel,\perp}= 8 \pi n_0 k_B T_{\parallel, \perp} / B_0^2$, temperature 
anisotropy, assumed $A= T_\perp / T_\parallel >1$, and plasma dispersion functions 
$Z_\kappa$ and $Z$ are explained in detail in Appendices A and B.

In the diffusion approximation time evolution of the electron distribution is described by 
the following quasilinear (QL) equation \citep{Yoon2017,LazarJGR2018}
\begin{align} \label{e2}
\frac{\partial f_e}{\partial t}&=\frac{i e^2}{4m_e^2 c^2}\int_{-\infty}^{\infty} 
\frac{dk}{k}\left[ \left(\omega^\ast-k v_\parallel\right)\frac{\partial}{v_\perp~\partial v_\perp}+ 
k\frac{\partial}{\partial v_\parallel}\right]\nonumber\\
&\times~\frac{ v_\perp^2~\delta B^2(k, \omega)}{\omega-kv_\parallel-|\Omega_e|}\left[ 
\left(\omega-k v_\parallel\right)\frac{\partial}{v_\perp~\partial v_\perp}+ k
\frac{\partial}{\partial v_\parallel}\right]f_e
\end{align}
combined with the wave kinetic equation 
\begin{equation} \label{e3}
\frac{\partial~\delta B^2(k)}{\partial t}=2 \gamma_k \delta B^2(k).
\end{equation}

Here $\delta B^2(k)$ is the wave energy density of whistler fluctuations, and
$\gamma_k$ is the instability growth rate obtained from Eq.~\eqref{e1}. The 
unstable fluctuations scatter the electrons reducing their anisotropy. 
Dynamical equations describing temperature components $T_{a, \perp, \parallel}$
(the second order moments of the velocity distribution) are readily obtained 
from Eq.~\eqref{e2}
\begin{align}
\frac{dT_{e, \perp}}{dt}&={m_e \over 2 k_B}\frac{d}{dt}\int d^3v ~ v_{\perp}^2~f_e \label{e4}\\
\frac{dT_{e,\parallel}}{dt}&= {m_e \over k_B}\frac{d}{dt}\int d^3v ~ v_{\parallel}^2~f_e \label{e5}.
\end{align}
Detailed expressions of Eqs. \eqref{e4} and \eqref{e5} are given in Appendix C. 

In order to outline the effects of suprathermal electrons here we compare the unstable whistlers 
obtained for a bi-Kappa distribution, with the unstable solutions obtained only for the 
corresponding bi-Maxwellian core, which is usually considered when suprathermals are ignored.  
Fig.~\ref{f1} illustrates linear and quasilinear solutions for three distinct cases corresponding 
to different (initial) plasma beta conditions $\beta_{c,\parallel} (0) = \beta_\parallel (0)
=$ 0.1, 0.5, 2.0, while the initial values of other parameters are the same, e.g., 
temperature anisotropy $A (0) = 4$ and power-index $\kappa=3$. The first two rows on top display 
linear growth rates $\gamma$, which are all systematically enhanced by suprathermal electrons (red 
lines), and wave-frequencies $\omega_r$ showing only minor variations, with a decrease of the 
most unstable frequencies. The stimulative effect obtained for the growth rates is 
further confirmed in the quasilinear evolution of the instability by the magnetic 
wave-power $U_B = \int \, dk \,\delta B^2(k)/B_0^2$ of the whistler fluctuations,
which shows the same systematic enhancement with an earlier initiation, a shorter growing time,
and markedly higher saturation levels reached in the presence of suprathermals. 
The immediate consequence of the more intense fluctuations is a faster and more effective 
relaxation of the temperature anisotropy $T_\perp/T_\parallel$. All these effects 
of suprathermal electrons become more pronounced with increasing the plasma beta parameter, 
from left to right columns in Fig.~\ref{f1}. The highest levels of the fluctuating magnetic 
power are proportional with the increase of $\beta_\parallel$. For an initial 
$\beta_\parallel (0) = 2$ the anisotropy drops rapidly and decreases asymptotically to very 
low values approaching stable states of isotropic temperatures.

Fig.~\ref{f2} displays temporal profiles of the full wave-number spectra of the magnetic 
power $\delta B^2(k)/B_0^2$, which show again an important increase in intensity in the 
presence of suprathermal electrons for all wave-numbers (upper panels). Magnetic power 
increases with increasing $\beta_\parallel$ (from left to right), and an earlier ignition 
of the instability is also confirmed in this case.

\begin{figure*}[h!]
   \centering
\includegraphics[width=0.31\textwidth]{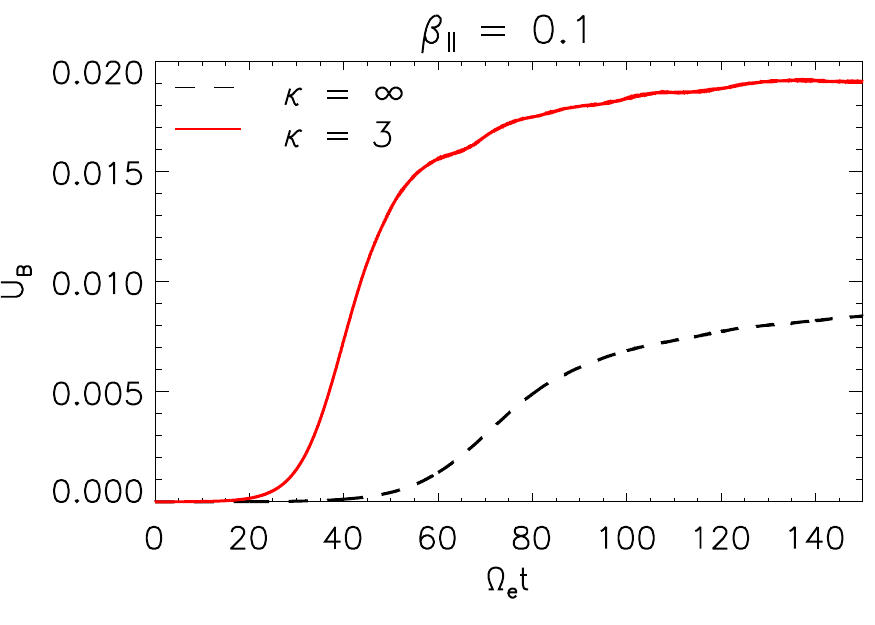} \includegraphics[width=0.31\textwidth]{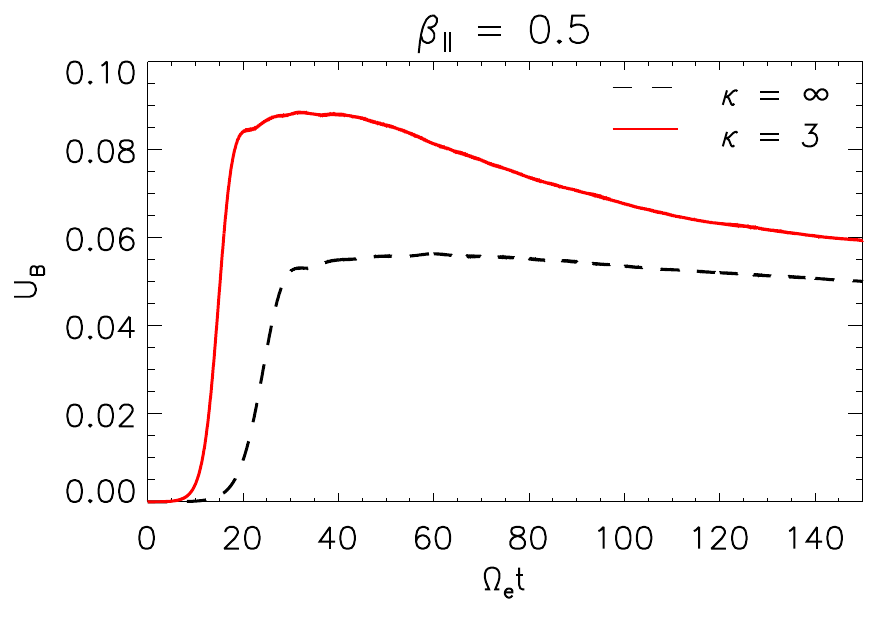}
\includegraphics[width=0.31\textwidth]{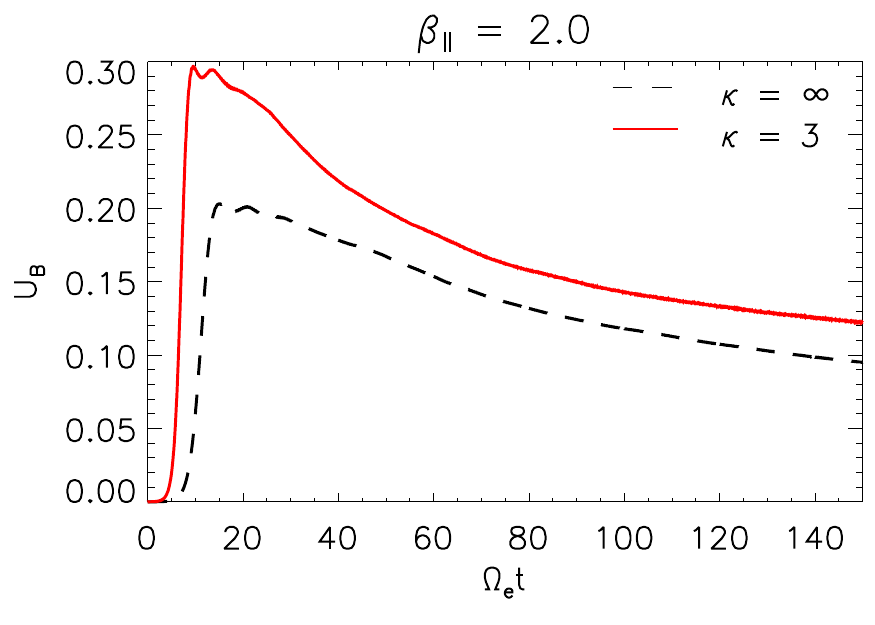}\\

\includegraphics[width=0.31\textwidth]{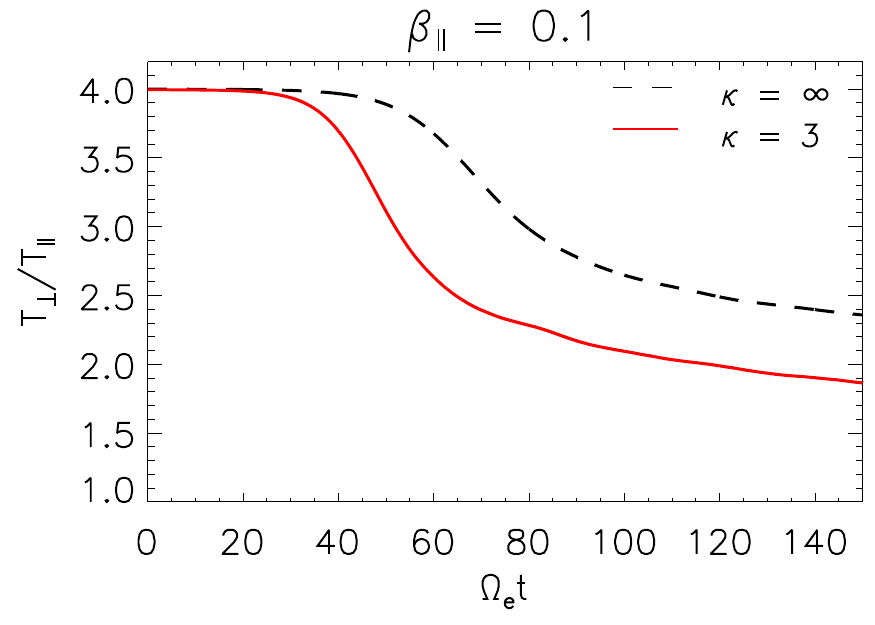} \includegraphics[width=0.31\textwidth]{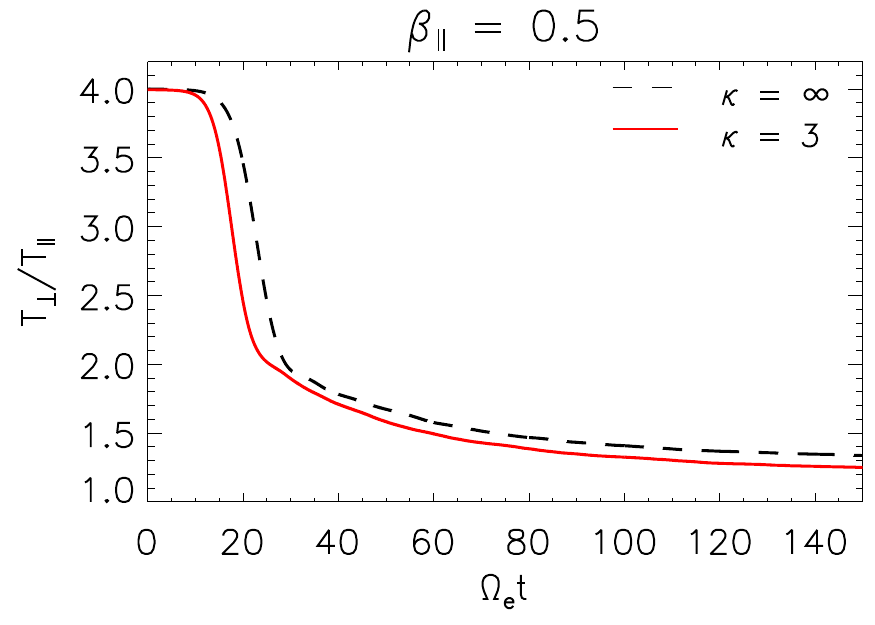}
\includegraphics[width=0.31\textwidth]{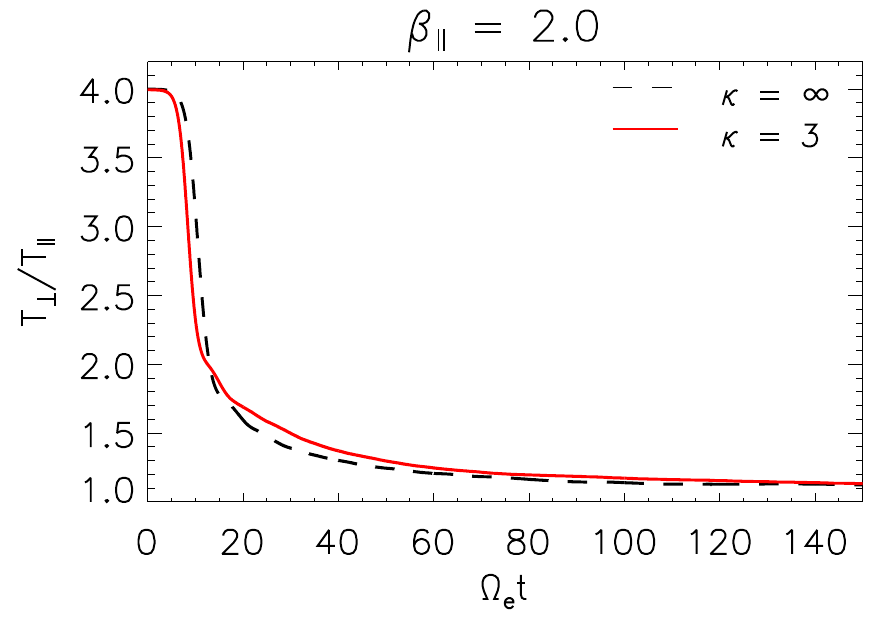}
\caption{Results from PIC simulations confirm a systematic stimulation of the magnetic power of 
whistler fluctuations (upper panels) and more efficient relaxation of temperature anisotropy 
in the presence of suprathermals, for the same three cases as in Fig.~\ref{f1}.} \label{f4}
\end{figure*}
\begin{figure*}[h!]
   \centering
\includegraphics[width=0.31\textwidth]{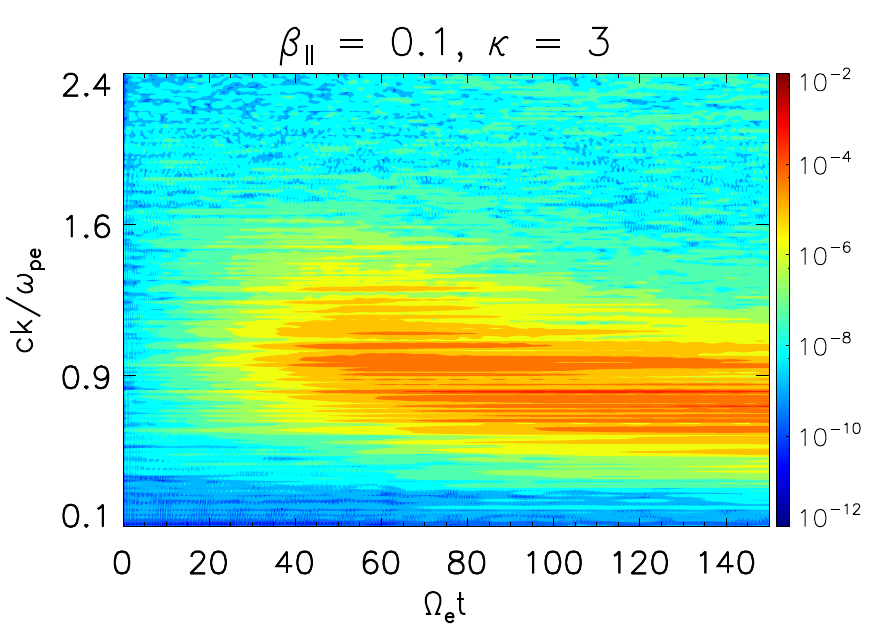} \includegraphics[width=0.31\textwidth]{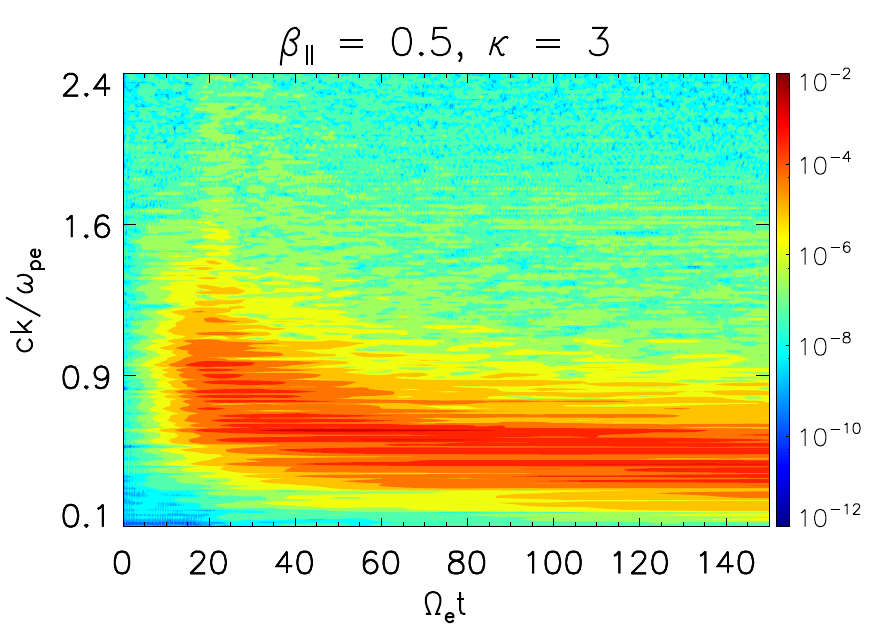}
\includegraphics[width=0.31\textwidth]{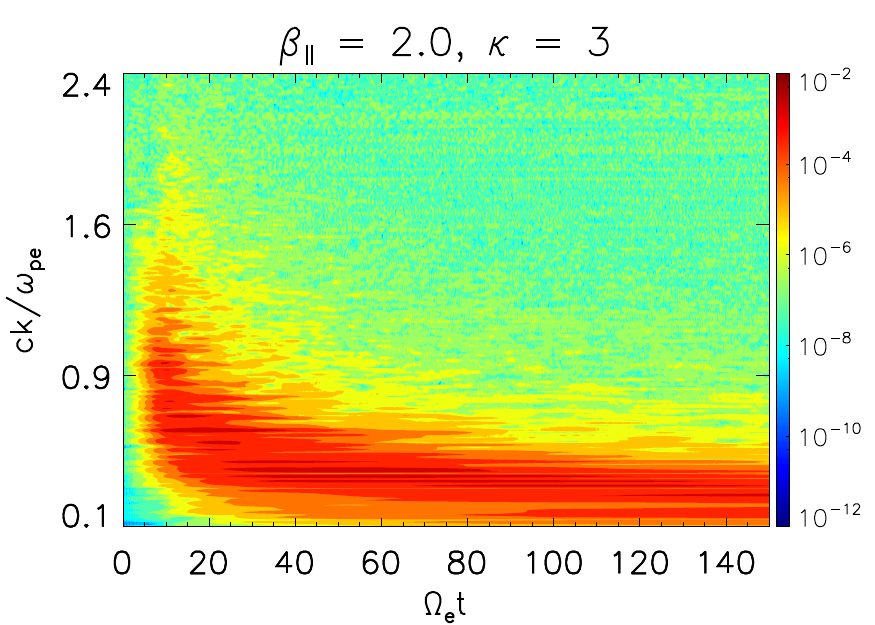}\\

\includegraphics[width=0.31\textwidth]{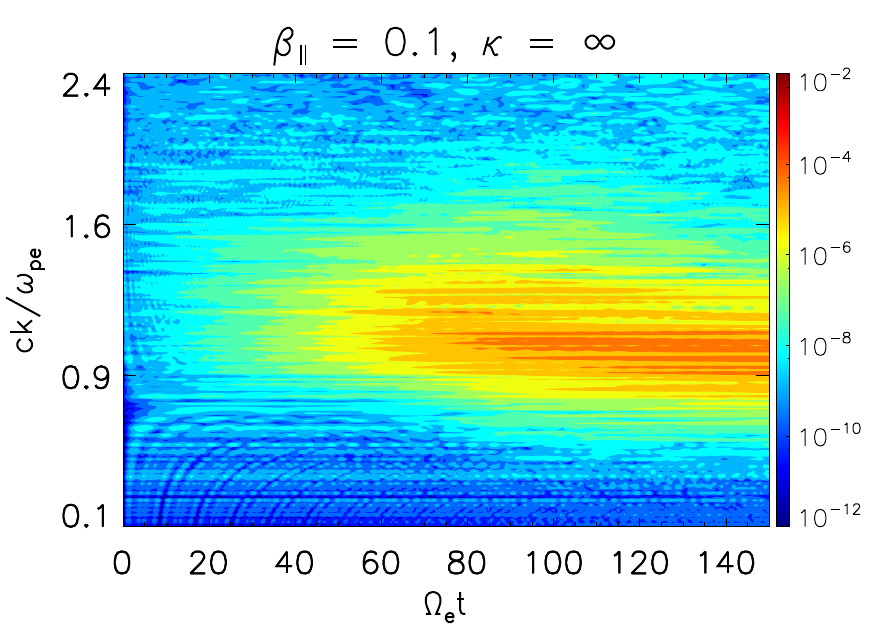} \includegraphics[width=0.31\textwidth]{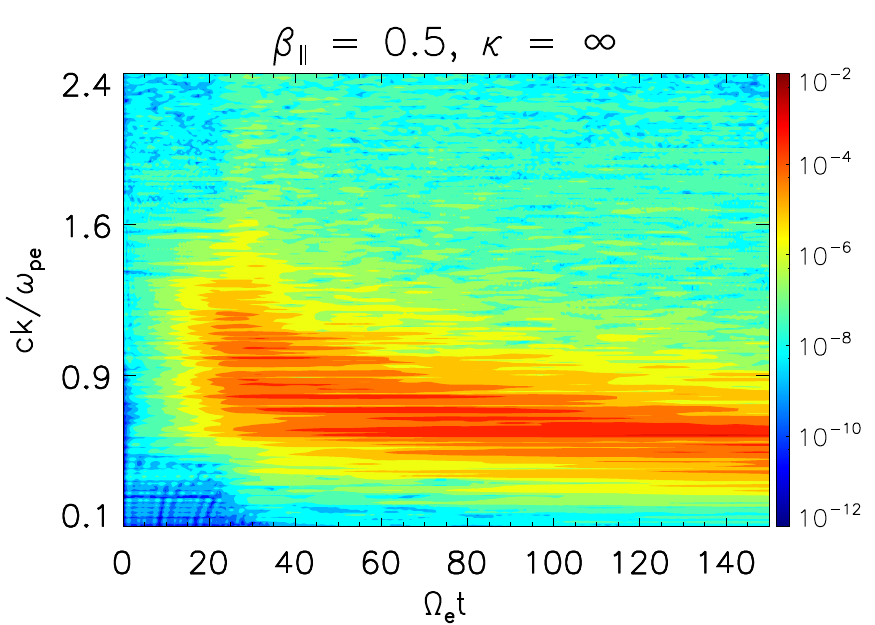}
\includegraphics[width=0.31\textwidth]{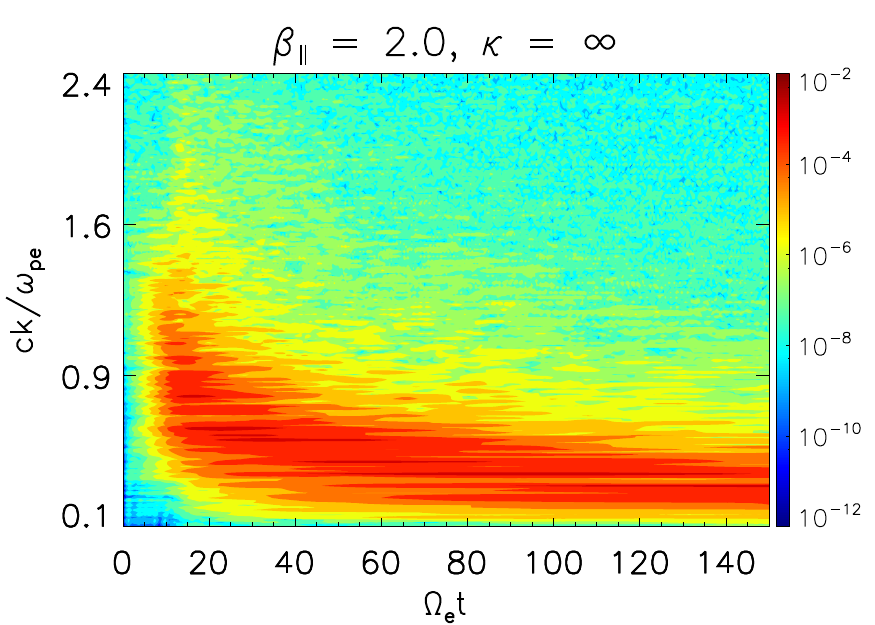}
\caption{Temporal evolutions for the wave-number spectra from PIC simulations confirming
the enhancement effect of suprathermals (upper panels) predicted by the theory in Fig.~\ref{f2}.
The color scale corresponds to the magnetic field power (see for details 
in the text).} \label{f5}
\end{figure*}

An even more comprehensive and detailed picture is offered by the anisotropy thresholds
plotted in Fig.~\ref{f3}, which describe the unstable solutions for a given value of the 
maximum growth rate $\gamma_{\rm max}$. Here we chose a low value $\gamma_{\rm max} = 
\gamma_m = 3 \times 10^{-3} |\Omega_e|$, approaching marginal stability and much lower 
than peaking values obtained in Fig.~\ref{f1}. The anisotropy thresholds predicted by 
the linear theory are fitted to
\begin{align}
A_e=1+\frac{s}{\beta_{\parallel}^{\alpha}}
\end{align}
with $(s, \alpha)= (0.18,0.52)$ for $\kappa=3$ (red curve) and $(s,\alpha)= 
(0.29, 0.49)$ for $\kappa \rightarrow \infty$ (dashed black curve). 
All dynamical paths derived from QL theory for the same three cases in Fig.~\ref{f1} 
converge to the same thresholds predicted by linear theory. The threshold values decrease 
under influence of suprathermal electrons (red curve), reconfirming their 
stimulative effect on whistler instability.

\begin{figure*}[ht]
   \centering
\includegraphics[width=0.31\textwidth]{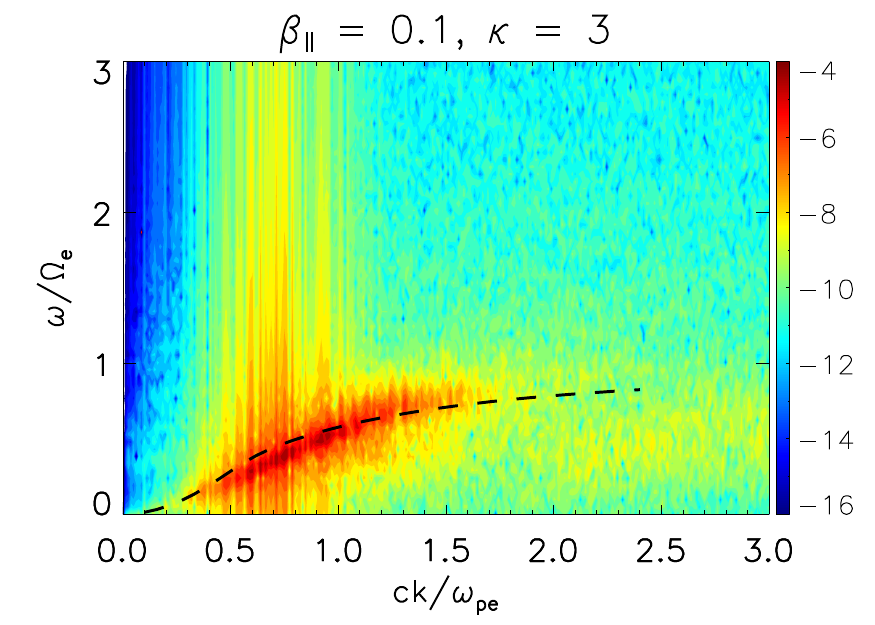} \includegraphics[width=0.31\textwidth]{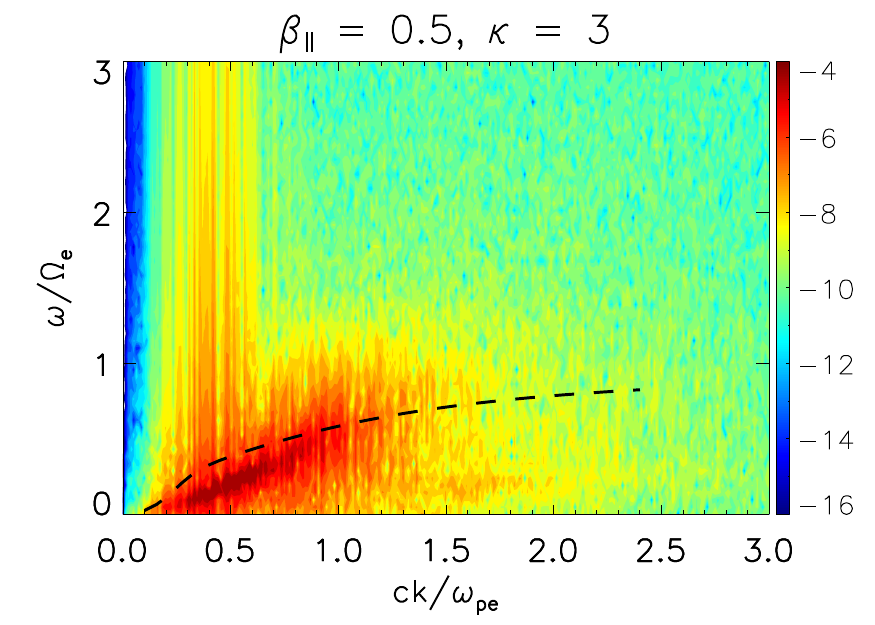}
\includegraphics[width=0.31\textwidth]{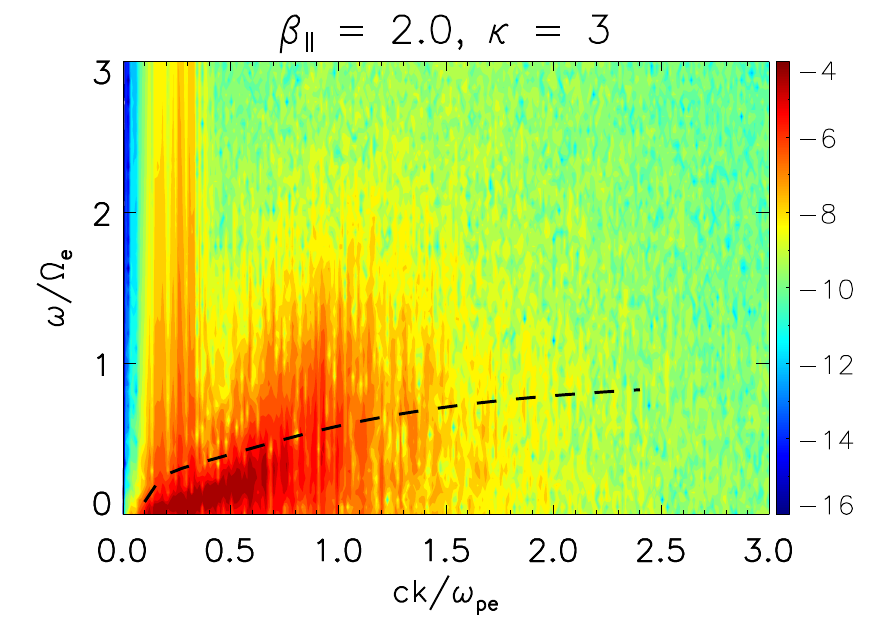}\\

\includegraphics[width=0.31\textwidth]{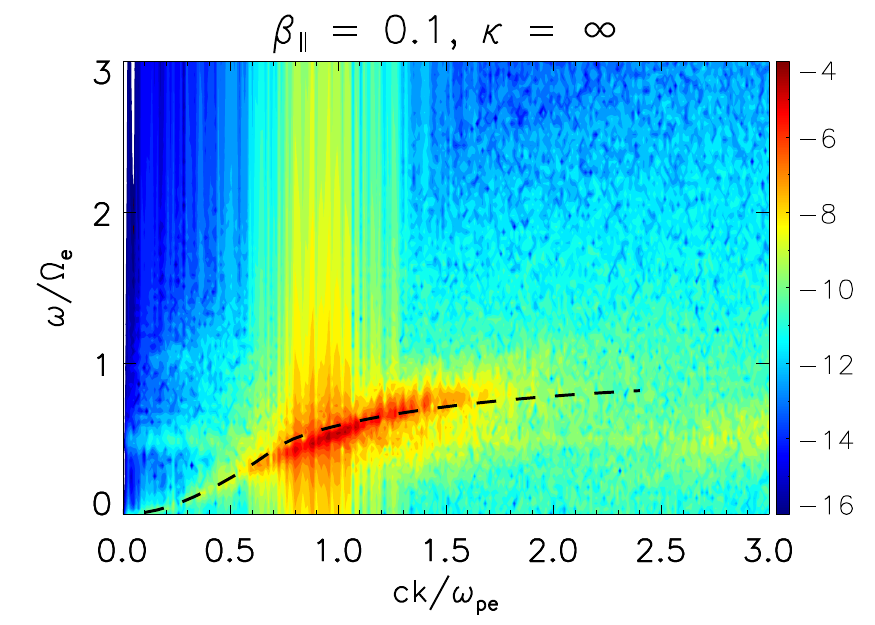} \includegraphics[width=0.31\textwidth]{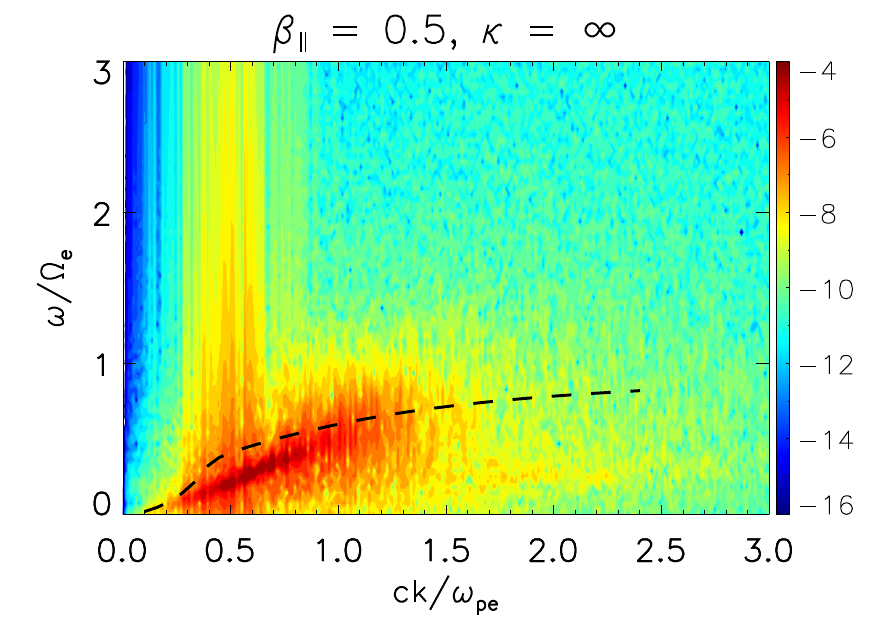}
\includegraphics[width=0.31\textwidth]{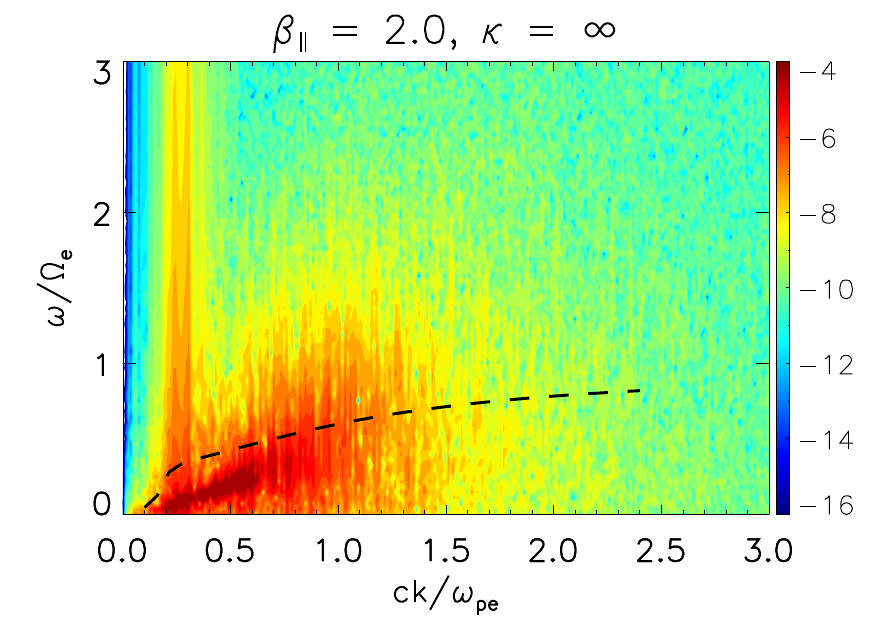}
\caption{Whistler fluctuations from PIC simulations resemble linear dispersion in the ($\omega-k$)-space 
and are enhanced in the presence of suprathermals (lower panels), and with increasing $\beta_\parallel$ (from left 
to right). The color scale corresponds to the magnetic field power (see for details 
in the text). } \label{f6}
\end{figure*}

\section{PIC simulations}

For further examination in numerical experiments we have used one-dimensional 
particle-in-cell (PIC) simulations, and the results show a very good agreement 
with QL theory. Let us first describe the plasma parameters and the setup used 
in the simulations. We assume a collisionless and homogeneous plasma of 
electrons (subscript $e$) and protons (subscript $p$) in the same conditions 
considered in theory, i.e., with the same values for the plasma beta parameters, 
a realistic mass ratio $m_p/m_e=1836$ and $\omega_{pe}/\Omega_e=20$ for the 
plasma to cyclotron frequency ratio. Electrons are described by a bi-Kappa 
distribution function with the same initial $\beta_{\parallel} =$ 0.1, 0.05, 2, 
corresponding to the three distinct cases considered in QL study, while protons 
are assumed Maxwellian and isotropic with $\beta_p = \beta_\parallel$. We used 
$n_x=2048$ grid points and $10000$ particles per grid cell, per species. The box 
size is $L_x=512\,c/\omega_{pe}$ in terms of the electron inertial length. The 
time step used is $\Delta t=0.01/\omega_{pe}$ and the simulation runs until 
$t_\text{max}= 150/\Omega_e$. To construct the power spectra as functions of 
the wave-frequency $\omega$ or the wave-number $k$ we have computed the Fourier-Laplace 
transforms of the magnetic field fluctuations in space and time.

Fig.~\ref{f4} illustrates the time variation of the total magnetic power $U_B$ 
(upper panels) and temperature anisotropy $T_\perp/T_\parallel$ (bottom panels) 
obtained from PIC simulations for the three cases shown in Fig.~\ref{f1}. The 
qualitative and quantitative agreement with the results from QL theory is obvious, 
confirming that suprathermal electrons stimulate not only the instability 
growth rates and the resulting magnetic power, but implicitly also the relaxation of 
temperature anisotropy. In Fig.~\ref{f5} we display the time evolution of 
the wave-number spectra of magnetic field fluctuations. The color scale corresponds to 
the magnetic field power, explicitly given by $|\text{FFT}_x(B_y-iB_z)|^2$, where 
$\text{FFT}_x$ is the fast Fourier transform along the spatial dimension. These 
spectra agree very well with the corresponding ones from QL theory in Fig.~\ref{f2}.
The saturation suggests a subsequent inverse cascade of the large-amplitude 
fluctuations towards lower wave-numbers, as already noticed by \cite{Kim2017} for 
Maxwellian plasmas. 

Finally, in Fig.~\ref{f6} we plot the total magnetic field power (integrated over the whole
interval of time) in $\omega-k$-space, enabling us to compare the trace of peaking intensities,
with the dispersion relation $\omega_r(k)$ derived from linear theory, see also \cite{Hughes2016}.
These plots are showing only the magnetic field fluctuations, as $|\text{FFT}_{(x,t)}(B_y-iB_z)|^2$
in the color bar, where the fast Fourier transform is taken along the spatial and temporal dimensions. 
Though this comparison is not rigorously motivated, since the plasma conditions 
change in time, in Fig.~\ref{f6} the pattern of peaking intensities resemble, especially 
for low-beta conditions ($\beta_\parallel = 0.1$), the linear dispersion (here shown 
with dashed lines) from Fig.~\ref{f1}. Notice again the enhancement of whistler fluctuations 
in the presence of suprathermals for all three cases (upper panels), which show the same
increase of intensities with increasing plasmas beta (from left to right). Unfortunately, the 
quasithermal noise is enhanced in the same measure, and where present (shown is 
only the electromagnetic component), it may determine significant deviations from theoretical 
dispersion (dashed lines). Only solutions obtained for low-beta conditions, e.g., $\beta_\parallel 
= 0.1$ are less affected by the quasithermal noise and approach better the 
linear dispersion, as also shown in \cite{Hughes2016} for bi-Maxwellian electrons
(see also their section V for more explanations on the effects of particle noise).



\section{Discussions and conclusions}

Investigated here is the whistler instability driven by a temperature anisotropy $T_\perp
> T_\parallel$ of electrons in conditions typically encountered in space plasmas, 
e.g.,  solar wind and planetary magnetospheres, where suprathermal particle populations 
are ubiquitous. Suprathermals enhance the high energy tails of the distributions, and 
overall, the observed velocity distributions are well described by the Kappa power-laws,
which are nearly Maxwellian at low energies and decrease as a power-law at higher energies.
As motivated in section~1, we assume the anisotropic electrons well described by a single 
bi-Kappa distribution function. Deviations from thermal equilibrium suggest an additional 
source of free energy in the presence of suprathermal electrons, that 
may be released through the effect of instabilities, as already suggested by linear theory 
\citep{Lazar2015Destabilizing,LazarMNRAS2017}.
In order to capture these mutual effects, here we have used a quasilinear (QL) kinetic 
approach and PIC simulations, which allowed us to characterize the long-term evolution of 
the instability under the influence of suprathermal electrons. 

To outline the effects of suprathermal electrons we have invoked a straightforward comparison 
of the observed bi-Kappa distribution with the quasi-thermal bi-Maxwellian core, which is 
solely considered when suprathermal (halo) populations are ignored. The same method
has already been applied in linear theory that confirms the expectations showing
an enhancement of the instability growth-rates in the presence of suprathermal electrons.
(and not only for whistler instability, as already shown in the Introduction). For whistlers 
enhanced growth-rates are plotted in Fig.~\ref{f1}, top line panels, for three distinct 
cases corresponding to $\beta_\parallel =$ 0.1, 0.5, 2. We have chosen these cases as 
representative for a large variety of space plasma conditions, ranging from low-beta 
plasmas in the outer corona or terrestrial magnetosphere, to a high-beta solar wind at 
large heliospheric distances. In addition, the results obtained here from QL theory and 
simulations are in perfect agreement, and show a systematic stimulation (no inhibition)
of the spectral and total magnetic power of whistler fluctuations. Moreover, a direct consequence 
of larger amplitude fluctuations is their back effect on the anisotropic distributions,
which show a faster and more efficient relaxation in the presence of suprathermals.
We have also shown that these effects markedly increase with increasing the plasma beta 
parameter. These results support the hypothesis that suprathermal populations are an 
important source of free energy, which stimulate the kinetic instabilities and implicitly,
the relaxation of the non-equilibrium collisionless plasmas from space. Moreover, our 
present results should also stimulate future studies on instabilities of different nature,
like firehose instabilities driven by an opposite anisotropy of temperature $T_\parallel >
T_\perp$, or beam-plasma instabilities with an influence of drifting suprathermals like 
the so-called electron strahl very often reported by the observations in the solar wind.

\begin{acknowledgements}
\textbf{Acknowledgements} The authors acknowledge support from the Katholieke 
Universiteit Leuven and Ruhr-University Bochum. These results were obtained 
in the framework of the projects SCHL 201/35-1 (DFG-German Research Foundation), 
GOA/2015-014 (KU Leuven), G0A2316N (FWO-Vlaanderen), and C 90347 (ESA Prodex 9). 
S.M. Shaaban would like to acknowledge the support by a Postdoctoral Fellowship 
(Grant No. 12Z6218N) of the Research Foundation Flanders (FWO-Belgium).
\end{acknowledgements}

\section*{Appendix A: Kappa vs. Maxwellian models}

In the presence of a suprathermal population the electrons are described by a bi-Kappa 
velocity distribution function \citep{Summers1991}
     \begin{align}
f_{\kappa} \left(v_{\parallel },v_{\perp }\right) = 
\frac{1}{\pi ^{3/2} u_{\perp }^{2}u_{\parallel}}&\frac{\Gamma\left(\kappa 
+1\right) }{\kappa^{3/2}\Gamma \left(\kappa -1/2\right) } \nonumber \\ 
& \times \left[ 1+\frac{v_{\parallel }^{2}}{\kappa u_{\parallel }^{2}}+
\frac{v_{\perp }^{2}}{\kappa u_{\perp }^{2}}\right] ^{-\kappa-1}\label{a1}
     \end{align}
which is normalized to unity $\int d^{3}v~f_{\kappa}=~1$ and is written in terms of 
normalization velocities $u_{\parallel, \perp}$ related to the components of kinetic 
temperature (for $\kappa_e >3/2$)
\begin{align}
T_{\kappa,\parallel} & =\frac{m}{k_B}\int d^3v v_\parallel^2 f_{\kappa}(v_\parallel, v_\perp) 
= \frac{\kappa}{\kappa-3/2}\frac{m_e u_{\parallel}^2 }{2 k_B}, \nonumber \\
T_{\kappa,\perp} & =\frac{m}{2 k_B}\int d^3v v_\perp^2 
f_{\kappa}(v_\parallel, v_\perp)=\frac{\kappa}{\kappa-3/2}\frac{m_e u_{\perp}^2}{2 k_B}.  
\label{a2}
\end{align}
Suprathermals enhance the high energy tails of the observed distributions, highly 
contrasting with the quasi-thermal core (subscript $c$) which is well described by 
a standard bi-Maxwellian model \citep{Lazar2015Destabilizing, LazarAA2016}
\begin{equation}
f_{c}\left(v_{\parallel },v_{\perp }\right) =\frac{1}{\pi ^{3/2}u_{\perp }^{2}
u_{\parallel}}\exp \left(-\frac{v_{\parallel }^{2}} {u_{\parallel }^{2}}-\frac{v_{\perp
}^{2}}{u_{\perp }^{2}}\right),  \label{a3}
\end{equation}
where $u_{\parallel, \perp}$ become thermal velocities corresponding to the temperature 
components
\begin{align} 
T_{c,\parallel} & =\frac{m}{k_B}\int d^3v v_\parallel^2
f_c (v_\parallel, v_\perp)=\frac{m u_{c,\parallel}^2}{2 k_B}, \nonumber \\
T_{c,\perp} & =\frac{m}{2 k_B}\int d^3v v_\perp^2
f_c(v_\parallel, v_\perp)=\frac{m u_{c, \perp}^2}{2 k_B}. \label{a4}
\end{align}
By comparison to the Maxwellian core, suprathermals also enhance the electron 
kinetic (energy) temperature \citep{Lazar2015Destabilizing}
\begin{align}
T_{\kappa,\parallel, \perp}=\frac{\kappa}{\kappa-3/2} \;T_{c,\parallel,\perp}
> T_{c,\parallel,\perp},\label{a5}
\end{align}  
and implicitly the plasma beta parameter%
\begin{align}
\beta_{\kappa,\parallel, \perp}=\frac{\kappa_e}{\kappa_e-3/2} \;\beta_{c,\parallel,\perp} 
> \beta_{c,\parallel,\perp} \equiv \beta_{\parallel,\perp}. \label{a6}
\end{align}  
The anisotropy does not depend on $\kappa$ and we can write generically
\begin{align}
A \equiv {T_{\perp} \over T_{\parallel} }= {T_{\kappa,\perp} \over T_{\kappa, \parallel}}=
{T_{c,\perp} \over T_{c, \parallel}}. \label{a7}
\end{align}
Playing only the role of a neutralizing background, the much heavier protons do 
not react to high-frequency modes, and, for simplicity, we can consider them 
isotropic and Maxwellian distributed with $\beta_p = \beta_\parallel$.

\section*{Appendix B: Plasma dispersion functions}

For the dispersion theory of Kappa distributed plasma we use the modified Kappa 
dispersion function \citep{Lazar2008}
\begin{align}
Z_{\kappa}\left(\xi_{\kappa}^{\pm }\right) & =\frac{1}{\pi ^{1/2}\kappa^{1/2}}
\frac{\Gamma\left(\kappa \right) }{\Gamma \left(\kappa-1/2\right) } \nonumber \\     
& \times \int_{-\infty }^{\infty} \frac{\left(1+x^{2}/\kappa \right) ^{-\kappa}}{x-
\xi_{\kappa}^{\pm }}dx, \; \Im \left(\xi _{\kappa}^{\pm }\right) >0.  \label{b1}
\end{align}
%
%
In the Maxwellian limit $\kappa \to \infty$ the plasma dispersion function 
takes the following standard form \citep{Fried1961}
\begin{equation}
Z\left(\xi^{\pm }\right) =\frac{1}{\pi ^{1/2}}\int_{-\infty}^{\infty}\frac{\exp\left(-x^{2}\right)}
{x-\xi^{\pm}}dt,\;\; \Im \left( \xi^{\pm }\right) >0,  \label{b2}
\end{equation}
which is used to describe the dispersion and stability of the electron core population,
when suprathermals are neglected. The proton term (subscript $p$) in Eq.~\eqref{e1} 
contains the same plasma dispersion function $Z$. 
%
%

\section*{Appendix C: Dynamical equations for the temperature components}

Time evolutions of the second order moments giving the components of electron 
temperature are obtained using \eqref{e2} in Eqs. \eqref{e4} and \eqref{e5} \citep{ShaabanApJ2019}
\begin{align}
\frac{dT_{\perp}}{dt} = & - \frac{e^2}{2m_e k_B c^2} \int_{-\infty}^{\infty} \frac{dk}{k^2} 
\langle~ \delta B^2(k)~\rangle \nonumber \\
&  \times \Bigg( \left(2A-1\right) \gamma + \text{Im}\left[{2i\gamma +\Omega_e \over ku_\parallel}~G \right] \Bigg), \label{c1} \\
\frac{dT_{\parallel}}{dt} = & \frac{e^2}{m_e k_B c^2} \int_{-\infty}^{\infty}\frac{dk}{k^2}
\langle~ \delta B^2(k) ~\rangle\nonumber \\ 
&\times \Bigg( A~\gamma  + \text{Im} \left[{\omega +\Omega_e \over ku_\parallel}~ G \right] \Bigg), \label{c2}
\end{align}
with 
\begin{align}
G &= \left[ A~\omega\pm\Omega_e\left(A-1\right)\right]Z_\alpha 
\left(\frac{\omega- |\Omega_e|}{k u_{\parallel}}\right), \label{c3}
\end{align}
where $Z_\alpha = Z$ for the Maxwellian core and $Z_\alpha = Z_\kappa$ for the observed Kappa distribution function.

\bibliographystyle{spr-mp-nameyear-cnd}
\bibliography{allpapers}

\end{document}